# The influence of nitrogen ion implantation on the microstructure and chemical composition of a thin layer on the biodegradable Zn-0.8Mg-0.2Sr substrate

Jan Pinc [1,*], Petr Vlčák[2], Miroslav Lebeda[1,2], Vilém Bartůněk[3], Vojtěch Smola[2], Marek Vronka[1], Jan Drahokoupil[1,2], Zdeněk Weiss[1], Petr Svora[1], Hana Lesáková[5], Kateřina Šindelářová[5], Orsolya Molnárová[1], Tomáš Horažďovský[2], Tomáš Studecký[3], Pavel Salvetr[3], Jiří Kubásek[5], Jaroslav Čapek[1] and Andrea Školáková[1,*]

[1] FZU - Institute of Physics of the Czech Academy of Sciences, Na Slovance 1999/2, Prague 8, 182 00, Czech Republic
[2] Department of Physics, Faculty of Mechanical Engineering, Czech Technical University in Prague
[3] Department of Inorganic Chemistry, University of Chemistry and Technology Prague, Technická 5, 166 28, Prague 6, Czech Republic
[4] COMTES FHT, a.s., Průmyslová 995, 334 41 Dobřany
[5] Institute of Metals and Corrosion Engineering, University of Chemistry and Technology, Technická 6, Prague 6, 166 28, Czech Republic
**\*** Corresponding authors: pinc@fzu.cz; skolakova@fzu.cz

**Abstract**
In this research, the influence of the N$^+$ ion implantation process on the microstructure of a biodegradable Zn-0.8Mg-0.2Sr alloy was investigated using various experimental techniques. Microscopic analysis revealed that a fluence of $17 \cdot 10^{17}$ ions/cm² resulted in the oversaturation of pure Zn and Mg$_2$Zn$_{11}$ surfaces, leading to the formation of nano/micro-porous layers up to 400 nm thick. The behavior of the Zn-0.8Mg-0.2Sr alloy was observed to be similar to that of the individual pure phases, albeit without the creation of pore structures. A limited formation of MgO and Mg$_3$N$_2$ was observed on the alloy surface, although the overall presence of Mg significantly increased from 0.8 to 15 wt.%. This increase was caused by the decomposition of the Mg$_2$Zn$_{11}$ phase during the process and subsequent diffusion of Mg toward the surface. The absence of Zn$_3$N$_2$ within the samples could be explained by the thermodynamic instability and low Zn-N affinity. Despite the absence of zinc nitride, GD-OES confirmed 10 at. % of nitrogen in the pure zinc, suggesting a possible accommodation of N atoms in the interstitial positions. This study points out to the complex nature of the process and highlights other promising directions for future research.

## 1. Introduction

Zinc and its alloys are considered to be members of biodegradable metals due to their ability to gradually degrade under physiological conditions, thereby the need for secondary surgeries to remove implants is eliminated [1]. The advantages and disadvantages of Zn-based materials concerning corrosion rate, mechanical performance, and cell/material interaction are well-documented in various publications [2-6]. Recently, there has been a significant increase in attention towards surface modification of these materials, primarily due to challenges in their interaction with tissue at the implant interface. These challenges are predominantly associated with a high sensitivity of the cells to zinc ion concentrations ranging between 40 and 80 µmol·L⁻¹, depending on the type of cell line [7]. Therefore, surface modifications of zinc are utilized to regulate corrosion rates [8], enhance cell adhesion [9], or prepare surfaces for drug delivery systems [10]. While numerous methods for modifying zinc surfaces have been proposed [11-13], such as laser modifications, atomic layer deposition, and chemical deposition, researchers continue to seek promising approaches to extend the usability of these materials.

Nitrogen ion implantation process has been known for a relatively long period of time; however, its usability was predominantly restricted to semiconductors [14], integrated circuits [15], fuel cells [16], or in a limited extend to permanent titanium implants [17]. This process involves doping the material surface with ions, inducing changes in surface properties through chemical and microstructural alterations. Compared to other surface modification methods, ion implantation offers several advantages, including precise process control, low processing temperatures, and enhanced adhesion of the modified layer to the substrate [18]. Additionally, the process can lead to changes in the chemical composition [17], microstructure [19], and roughness of the implanted material [20] to achieve desired characteristics for specific applications. These advantages point out ion implantation as a perspective method for creating homogeneous, functional layers in complex biodegradable systems.

According to the literature, only a limited group of zinc compounds, primarily ZnO, have been implanted using nitrogen (N) ions [21-23]. This limitation stems from the poor stability of $Zn_3N_2$ due to chemical reaction with air humidity, restricting the use of these materials. On the contrary, the poor stability of the layer is not an obstacle within the biodegradable materials. Besides, the chemical reaction can positively affect the healing process through pH regulation in the vicinity of the implant during inflammatory processes [24, 25]. However, the advantages of the ion implantation process extend beyond changes in chemical composition. Another notable benefit is the ability of the process to create nanoporous structures on the material's surface serving as a potential base material for drug delivery systems. Despite the formation of nanoporous structure not being observed within Zn materials yet, examples of such behavior can be seen in the case of implanted titanium. Vlcak et al. [26] revealed that the oversaturation of the titanium surface by $N^+$ ions leads to the formation of different nanopore structures dependent on the crystallographic orientation of the grains. In addition, Minagar et al. [27] described the enhanced proliferation of the cells on the nanostructured surface of titanium. Based on these observations, a positive effect of the $N^+$ ion implantation on the performance of biodegradable Zn materials can be expected.

According to our best knowledge, this is the first publication regarding the issue with a focus on potential usage in Zn biodegradable implants applications. This alloy was specifically selected for its sufficient mechanical performance [28] and well-documented corrosion behavior [29], which will makes it suitable for future comparisons between implanted and non-implanted materials. Additionally, the description of the process highlights the advantages of implanted surfaces compared to unmodified materials and suggests other directions for research in this field.

**2. Experimental**

In this study, forged Zn-0.8Mg-0.2Sr alloy was used as an initial material. The material was prepared from the pure elements containing Zn (99.995 wt. %), Mg (99.95 wt. %) and Sr (99.9 wt. %) in the desired weight ratio. Firstly, zinc was melted at 520 °C in the resistance furnace without the presence of the protective atmosphere. Secondly, pure magnesium was added into melt and the mixture was stirred using graphite rod to homogenize the composition of the melt. Lastly, the Sr wrapped into Zn foil (due to avoid Sr oxidation) was added and the mixture was left in the furnace for another 10 min. After the homogenization, the melt was poured into copper mold with a diameter of 70 mm and cooled down on the air. Subsequently, the ingot was annealed at 350 °C for 24h and quenched into water to homogenize its phase composition. To remove possible contamination concerning the Cu mold, homogenized samples were turned with minimal removal of the material equal to 1 mm. A hydraulic forging press was utilized to convert the ingot into a test sample. A cylindrical ingot, measuring 68 mm in diameter and 110 mm in length, was subjected to multiple forging operations to attain a uniform microstructure. The resulted dimensions of the square bar were 30 mm on each side. To maintain the material's plasticity, the piece was preheated and reheated to 220 °C. The final forging temperature was controlled to not fall below 170 °C. Die temperatures were maintained within a range of 170 °C to 230 °C. Temperature monitoring was conducted using a calibrated thermographic camera and a contact thermocouple, ensuring stable temperature during process. The upsetting process generated significant heat, with the highest observed surface temperature reaching 245 °C. The degree of deformation, quantified as the total relative compression, was confirmed to be 30 %. Forged bar was milled to remove damaged surface and cut to samples with 10x10x3 mm dimensions. Pure $Mg_2Zn_{11}$ phase was prepared in the resistance furnace using pure metals under protective Ar atmosphere. Composition of prepared phase was determined using X-ray diffraction as 96 wt. % of $Mg_2Zn_{11}$ and 4 % of Zn. All samples were polished using diamond suspension (1 µm) and Etosil E suspension ($SiO_2$, 0.06 µm) and subsequently used for the nitrogen implantation.

The samples were implanted with 90 keV nitrogen ions at a fluence of $1.7 \cdot 10^{18}$ cm$^{-2}$ in the direction perpendicular to axis of forging (PD). The other directions, parallel with axis of forging (FD), were not affected by the process. The ion beam current was measured using a Faraday cup, and the temperature of the sample was measured by a thermocouple located in the sample holder. The average ion beam current density was kept at 3 µA, and the work pressure during implantation varied between $1\text{-}3 \cdot 10^{-5}$ mbar. The temperature of the samples during implantation did not exceed 300 °C. Process conditions were selected based on previous experiments concerning the implantation of titanium samples. Implanted samples were stored in desiccator to prevent the changes in the chemical composition caused by the sensitive nature of the zinc nitride compound to air moisture.

Transport of Ions in Matter (TRIM) simulations were conducted utilizing the Stopping and Range of Ions in Matter (SRIM) 2013 package [30]. The preset parameters were employed for displacement energy, surface binding energy, and lattice binding energy specific to each element (Zn, Mg). The densities used for the Zn and $Mg_2Zn_{11}$ phases were set as $6.58 \cdot 10^{22}$ and $6.29 \cdot 10^{22}$ atoms·cm$^{-3}$, respectively [31, 32]. To ensure reasonably large statistic, a sample size of one million implanted N ions was applied with the 'Detailed Calculation with full Damage Cascades' mode to also obtain sputtering yields. The depth distributions for the implanted N were computed separately at two ion energy levels, 45 keV and 90 keV. The profiles derived from both scenarios were subsequently summed, with weighting factors of 0.824 for the 45 keV profile and 0.176 for the 90 keV profile. This weighting reflects the composition of the experimental beam (~70 % $N^{2+}$ and 30 % $N^+$).

TRIDYN [33] calculations were carried out in accordance with the software's recommended settings for bulk binding energy, relocation threshold energy, and surface binding energy for the constituent elements (Zn, Mg, N). The atomic density of Zn, acting as the principal component in both the Zn and $Mg_2Zn_{11}$ phases, was set as $6.58 \cdot 10^{22}$ atoms·cm$^{-3}$. The density for the non-principal Mg component was determined according to the equation (1) as $6.43 \cdot 10^{22}$ atoms·cm$^{-3}$, based on the density of $Mg_2Zn_{11}$ with $6.29 \cdot 10^{22}$ atoms·cm$^{-3}$. Parameter for the atomic density of implanted N (artificial density used to correctly reproduce the density of the possible resulted phase) was calculated considering the phase of $Zn_3N_2$ with $8.36 \cdot 10^{22}$ atoms·cm$^{-3}$ [34], yielding $14.06 \cdot 10^{22}$ atoms·cm$^{-3}$.

$$DNS_B = \left( \frac{n+m}{m} \frac{1}{DNS_{A_nB_m}} - \frac{n}{m} \frac{1}{DNS_A} \right)^{-1} \qquad (1)$$

where n, m are the subscripts in $A_nB_m$ compound, $DNS_B$ represents the density of the non-principal element B, $DNS_{AnBm}$ refers to the overall density of the $A_nB_m$, and $DNS_A$ is the density of the principal component A.

The samples were characterized using an FEI 3D Quanta 3D field-emission-gun DualBeam scanning electron microscope (SEM) equipped with an electron backscatter diffraction (EBSD) detector TSL/EDAX Hikari. The micrographs were captured in the backscattered electrons (BSE) regime at accelerating voltages of 10 kV, a working distance (WD) of 10 mm, and in a high vacuum mode. Thin foils for transmission electron microscopy (TEM) observations were prepared using the focused ion beam (FIB) technique. TEM and scanning transmission electron microscopy (STEM) images were captured by FEI Tecnai TF20 X-twin field emission gun transmission electron microscope operated at 200 kV equipped with an EDS detector. For STEM, Z-contrast imaging was performed using a high-angle annular dark field (STEM–HAADF) detector. Diffraction patterns were evaluated using the program CrysTBox [35].

Concentration-depth profiles of the elements were established by glow discharge optical emission spectroscopy (GD-OES) [36] using the GDA750HR spectrometer (Spectruma GmbH., Germany). A 2.5 mm internal anode Grimm-type spectral source was used and operated with a DC discharge in argon at 850 V/15 mA. Quantification of the profiles was based on a sputter rate-corrected calibration [36], involving all the elements analyzed.

The X-ray diffraction patterns were measured on Empyrian diffractometer from PANalytical company equipped with Cu tube ($\lambda$ = 1,54056 Å). The parallel beam geometry (parallel beam mirror in primary beam and parallel plate collimator with angle acceptance equal to 0.18° in diffracted beam) with incident angle equal to 0.5° was used. This low angle of incidence was used to reduce penetration depth. The effective penetration depth (63% of diffracted intensity) is for used settings equal to 117 nm in Zn (with density 7.14 g·cm$^{-3}$) or 770 nm in Mg (with density 1.738 g·cm$^{-3}$). The measurement from 10 to 100 ° [2Θ] takes approximately 6 h.

Atomic force microscopy (AFM) measurements were performed using polished and implanted samples. Within each sample, the quadratic mean (Rq) and arithmetic average roughness (Ra) were determined from minimum of six different locations. The measured areas were 100 x 100 µm. For measurement, the AFM NanoWizard3 instrument (JPK, Berlin, Germany) was used. The contact imaging mode in the air was used with a scanning speed set to 1 Hz, Setpoint to 1.30 V, and a resolution of 256 pixels. We used a CONTV-A cantilever (Bruker probe with Al reflective coating) with spring constant 0.2 N·m$^{-1}$, and frequency 13 kHz. To adjust for sample tilt, all topography images of the samples were flattened using the plane fitting option in the JPK data processing software. For illustration of thermodynamic stability in three phase systems Mg-Zn-N$_2$ and Mg-N$_2$-O$_2$, the FactSage 8.2 program was used [37]. Thermodynamic data were fetched from FactPS - FACT pure substance database (2022), containing data of both Zn$_3$N$_2$ and Mg$_3$N$_2$ compounds.

## 3. Results

**Process simulations**

To obtain the expected shapes of the implanted N depth distributions and to get insights into the experimentally observed profiles, we have performed Monte Carlo simulations within TRIM and TRIDYN code.

TRIM simulations revealed the nitrogen depth profiles under the assumption that the target's densities remain constant, and no sputtering occurs during the implantation. Under these conditions, the predicted N concentration maximum exceeded 70 at. % in Zn and Mg$_2$Zn$_{11}$ targets (**Fig. 1a**), significantly higher than what was observed experimentally. This discrepancy could suggest that a substantial amount of the implanted N escapes due to the formed oversaturation defects, such as cracks and pores. This is in good agreement with the results obtained within SEM and TEM observations (**Figs. 3, 4**).

In general, it is possible in a postprocessing of the TRIM data to explicitly calculate the sputtered depth based on the sputtering yield and the total implanted fluence. However, for such a high fluence as $18 \cdot 10^{17}$ ions/cm² and for the sputtering yield of 2.53 of Zn, the computation indicates a sputtered depth of approximately 8002 Å. For the Mg$_2$Zn$_{11}$, with a sputtering yield around 2.17, the calculation yields 6268 Å. These values are unrealistic as they surpass the depth range (around ~2500 Å) of the N ions across the target. Thus, for our experimental conditions, this effect cannot be incorporated in TRIM results.

To account for the sputtering and dynamic modification of the target composition due to the implanted ions, TRIDYN simulations were carried out. We have explored two specific scenarios: one without a predefined maximum concentration limit on the implanted N ions, allowing comparison with TRIM results, and another with a 15 at. % limit to align more closely with experimentally observed data. The aforementioned curves regarding the TRIDYN simulations are shown in **Fig. 1b**.

In the scenario without the limit, the peak concentration of N ions slightly falls below 60 N at.%. The peak itself is somewhat indistinct, with the concentration experiencing a gradual decline with the depth after a slight rise from the surface. The maximum observed concentration is about 10% lower and about 400 – 500 Å closer to the surface compared to the results obtained by TRIM simulations, highlighting the additional effects accounted for by TRIDYN. When the concentration limit of 15 at. % was imposed, the N concentration remained constant from the surface to a depth of approximately 850 Å, beyond which it began to decrease. This is similar to a trend observed in TEM analysis, showing good correspondence with the experimental data. Additionally, a distinct bump in concentration is observed at a depth of around 1100 Å, which can be attributed to the subset of single implanted ions with an initial energy of 90 keV.

When comparing the profile shapes of Zn and $Mg_2Zn_{11}$ computed by TRIDYN simulations, the Zn target exhibits slightly lower N concentrations across the entire depth range compared to $Mg_2Zn_{11}$. For the case of an imposed limit, the N concentration begins to decline at shallower depths. This behavior is caused by the higher sputtering yield for Zn compared to $Mg_2Zn_{11}$. Due to the specific characteristics of the utilized Monte Carlo codes, TRIDYN provided results that are more consistent with experimental data compared to TRIM. However, this outcome is expected, given that TRIM typically provides accurate results for low implantation fluences when the concentration remains below a few atomic percent. In such cases, the target density undergoes minimal change, and sputtering effect is not of significant importance. The observations from TRIM on our data illustrates the importance of utilizing a code like TRIDYN for applications involving high implantation fluences.

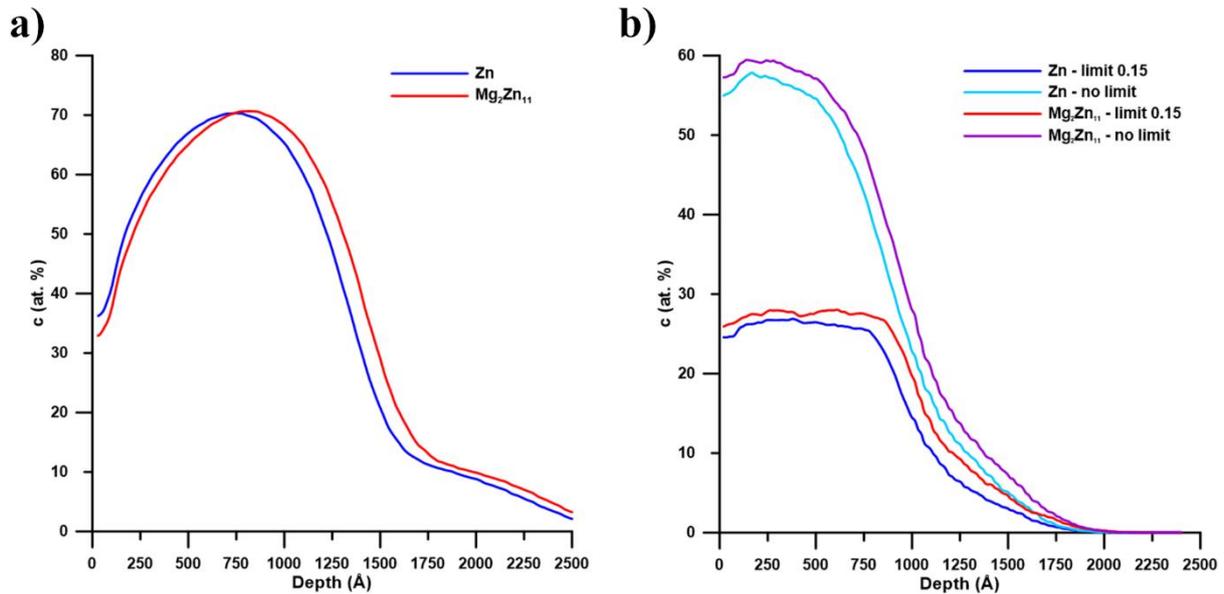

**Fig. 1:** The curves representing N concentration-depth relation determined using a) TRIM and b) TRIDYN simulations

**Initial material**

To reliably describe the process of $N^+$ ion implantation into the surface of the Zn-0.8Mg-0.2Sr alloy, it is necessary to characterize the microstructure, phase composition, and distribution of secondary phases in the initial material. The results pd SEM-EDX analyses giving this information are shown in **Fig. 2**. The microstructures of the as-cast and annealed Zn-0.8Mg-0.2Sr alloys have already been described in our previous studies [28, 38]. Based on that, only a brief description focusing on the forged alloy will be presented within the results section. The as-cast state of the alloy consisted of Zn dendrites surrounded by intermetallic regions containing eutectic systems Zn-$MgZn_2$, Zn-$Mg_2Zn_{11}$, and particles of the $SrZn_{13}$ phase. Homogenization of the phase composition is achieved through subsequent annealing during which the "metastable" $MgZn_2$ phase transforms into the "stable" $Mg_2Zn_{11}$ phase. Additionally, the interconnected net of eutectic regions is disrupted into particles of the $Mg_2Zn_{11}$ phase, with a size of tenths of micrometers during annealing. As observed in **Fig. 2a** and **b**, no changes in the phase composition were observed after forging. Clusters of relatively fine intermetallic particles were formed in the microstructure (**Fig. 2a**), arranged into lines perpendicular to forging. Changes in the arrangement of the secondary phases also indicate possible elongation of the zinc matrix grains. In the forged material, the average grain size of the zinc matrix reached approximately 30 µm in the FD, while in the PD, it reached an average of 20 µm. Finally, the depth profiles obtained by GD-OES are shown in **Fig. 2c**. Based on these results, a thin layer of zinc oxide can be expected on the material surface, with a gradual normalization of the composition with the depth of analysis. Additionally, a low content of nitrogen and hydrogen can be observed in **Fig. 2c**. Their presence, along with a portion of oxygen, could be attributed to the degassing of the inner parts of the discharge lamp during analysis.

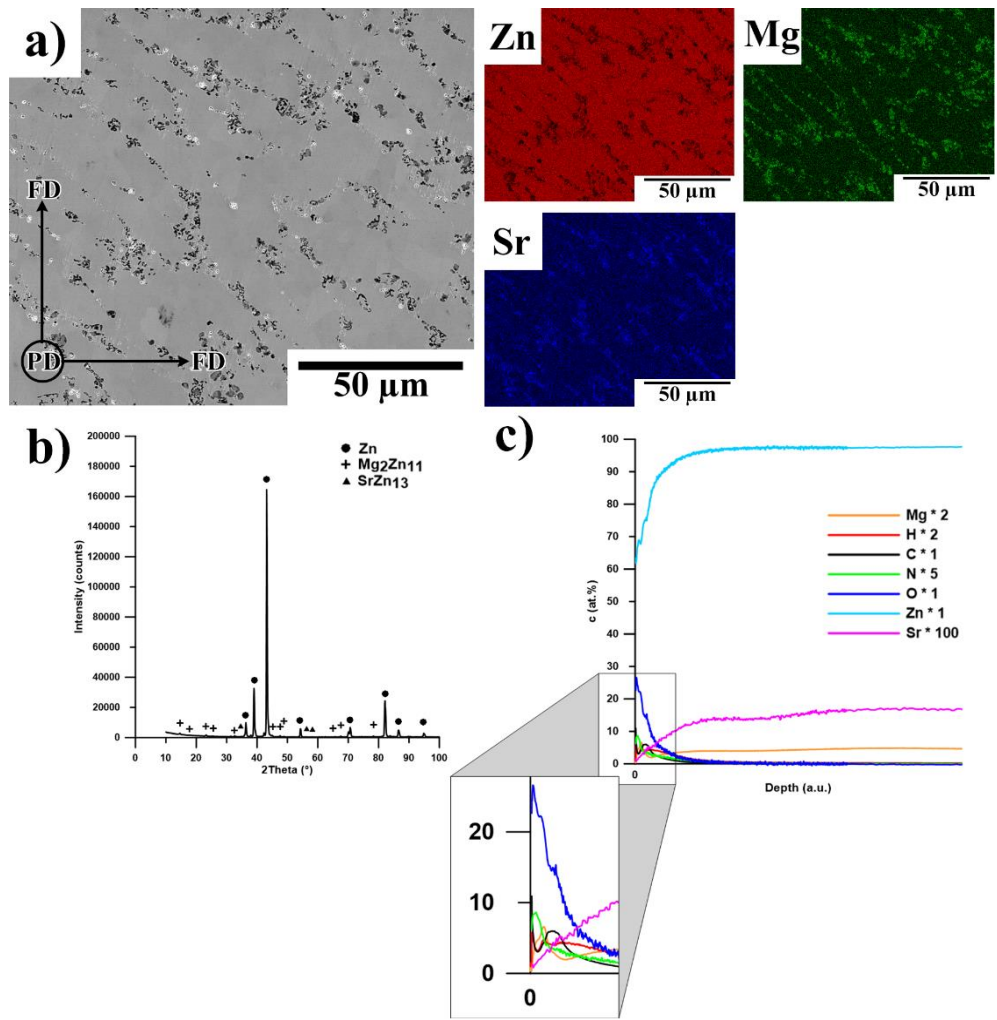

**Fig. 2:** a) Microstructure of the forged Zn-0.8Mg-0.2Sr sample with EDS element distribution maps; b) XRD pattern representing phase composition of the forged sample; c) depth profiles of the individual elements determined using GD-OES.

**Characterization of the samples after nitrogen implantation**

In addition to the description of the initial material, ion implantation was also performed on pure Zn and $Mg_2Zn_{11}$ to make the assumptions concerning the influence of the process on the Zn-0.8Mg-0.2Sr alloy even more precise. The resulting microstructures of all mentioned samples after implantation are shown in **Fig. 3**. As can be seen in **Fig. 3a**, small pores ranging in size from 40 to 600 nm were observed on the Zn surface with a homogeneous distribution all over the analyzed surface. On the contrary, the surface of the $Mg_2Zn_{11}$ phase contained cracks and a small number of holes with a diameter of 10 µm after implantation (**Fig. 3b**). Additionally, the amount of Zn increased in the cracks (see supplementary materials), suggesting a change of Zn-Mg ratio underneath the surface layer. As can be seen from **Fig. 3c**, the microstructure of the implanted Zn-0.8Mg-0.2Sr alloy significantly differs compared to initial Zn-0.8Mg-0.2Sr material or implanted $Mg_2Zn_{11}$ phase and Zn. The microstructure of the implanted alloy can be separated into two different areas represented by bright and dark areas in **Fig. 3c**. Bright areas were composed of Zn without a visible presence of pores as in the case of implanted Zn (**Fig. 3a**). Moreover, the presence of small structures with a round shape was observed on the surface of these areas as well. On the contrary, the dark areas were composed predominantly of Mg with minor concentration of Zn. Furthermore, cracks in the dark areas were observed approximately in the middle of these areas. One of the most important facts is that the ratio between

Zn and Mg significantly changed on the material surface in favor of Mg. More precisely, the content of Mg in terms of area increased by 15 wt. %.

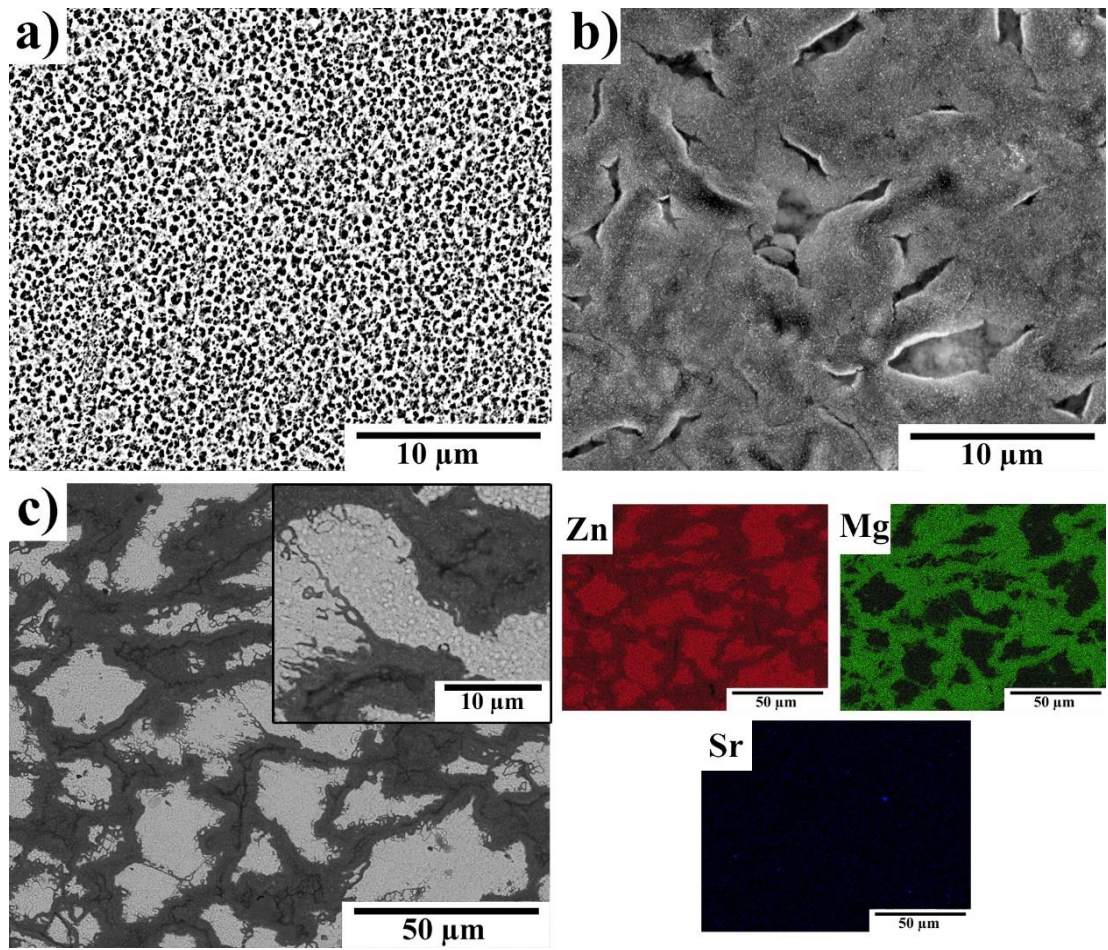

**Fig. 3:** SEM micrographs of implanted a) pure Zn, b) $Mg_2Zn_{11}$ phase and c) Zn-0.8Mg-0.2Sr alloy with elemental maps

To reveal the depth of implanted N ions and the character of the bond between N and the substrate, elemental profiles and electron diffraction were conducted on all three implanted samples. Results regarding the electron diffraction are shown in **Fig. 4**. In the case of implanted Zn (**Fig. 4a**), a pore structure was formed on the surface with a relatively homogeneous thickness of 270 nm. Additionally, the presence of N (up to 19 at. %) and O (up to 10 at. %) was detected in the aforementioned layer. However, the presence of ZnO or $Zn_3N_2$ was not confirmed by electron diffraction. The analysis of the implanted $Mg_2Zn_{11}$ phase revealed a more complex character of the layer compared to implanted pure Zn. It is evident from **Fig. 4b**, that the material consisted of three separated layers. The first one, closest to surface, was comprised of MgO and $Mg_3N_2$, and this layer was divided into two parts by pores occurring in the middle of the layer. Both phases formed a layer with a thickness of up to 400 nm. The second layer was composed of pure Zn with slightly increased content of O (< 8 at. %) and minimal content of N (< 3 at. %). The thickness of this layer reached approximately 700 nm. Beneath the second layer, the original $Mg_2Zn_{11}$ substrate was found. In the case of the alloy, two different lamellae were prepared from the bright (**Fig. 4c**) and dark areas (**Fig. 4d**) visible in **Fig. 3c**. In the case of bright area, the lamella only consisted of Zn, without the presence of other Mg- or Sr-based compounds. Additionally, a porous layer with a thickness close to 200 nm and a thin diffusion layer of Zn were confirmed, which is in good agreement with the results obtained for implanted pure Zn (**Fig. 4a**). Conversely, the second lamellae from dark area (**Fig. 4d**) possessed a similar layer arrangement as in the case of the implanted $Mg_2Zn_{11}$ phase, with small deviations in layer thickness and N concentrations.

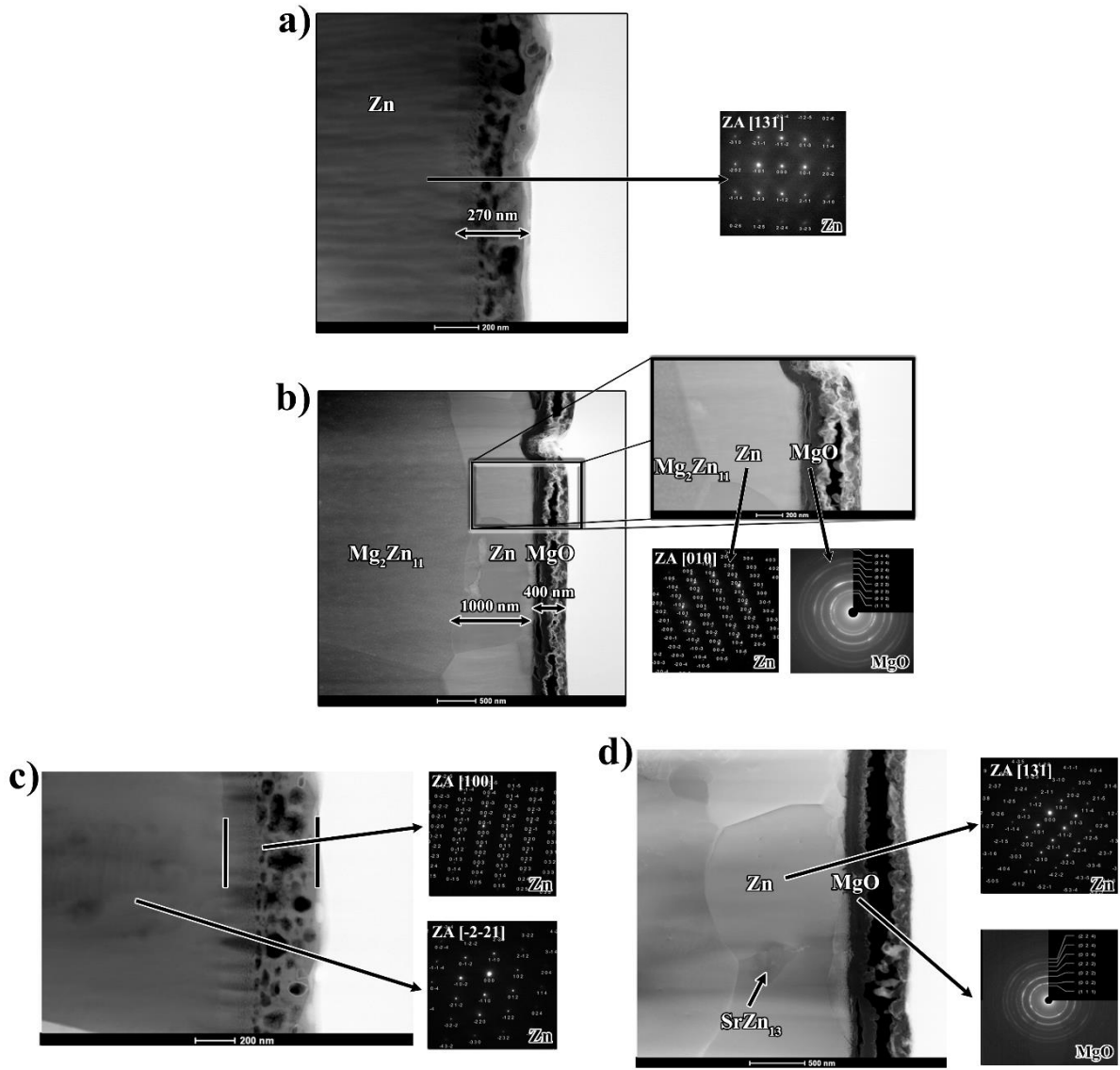

**Fig. 4:** STEM images and electron diffractions of implanted a) pure Zn, b) $Mg_2Zn_{11}$ phase, and Zn-0.8Mg-0.2Sr alloy in the areas containing majority of c) Zn and d) Mg on the surface of lamellae

Profile of the chemical composition was determined using GD-OES, and the results concerning implanted pure Zn, $Mg_2Zn_{11}$ phase, and Zn-0.8Mg-0.2Sr alloy immediately after implantation and after 3 days are shown in **Fig. 5**. In the case of the initial material (**Fig. 1c**), the maximum concentration of nitrogen reached approximately 2 at. %, suggesting the value as a potential error for implanted alloy. In the case of pure Zn (**Fig. 5a**), the nitrogen content reached approximately 10 at. % with an immediate decrease in concentration with depth. On the contrary, the concentration of nitrogen in the $Mg_2Zn_{11}$ phase was significantly higher with a rapid decrease in concentration in three steps (**Fig. 5b**). This higher concentration suggests the ability of the phase to accumulate nitrogen in the structure. The content of oxygen is also highly important due to possible chemical reactions connected with the transformation of nitrides into oxides. The oxygen concentration reached 17 at. % for pure Zn and 58 at. % for the $Mg_2Zn_{11}$ phase, respectively. In addition to implanted pure Zn and $Mg_2Zn_{11}$ phase, the profile was also measured for the implanted Zn-0.8Mg-0.2Sr alloy, together with the time stability (3 days) of the nitrides. As can be seen in **Fig. 5c** and **d**, the value of N and O concentrations were very similar in both cases. The nitrogen content reached approximately 7 at. % and oxygen 40 at. %, respectively. Despite the changes in nitrogen and oxygen concentrations, the increase of Mg on the surface of the alloy was visible as well. This is in good agreement with the SEM observations. Besides, the concentration of Mg gradually decreased to almost 0 at. % followed by its new growth stabilized afterward with depth.

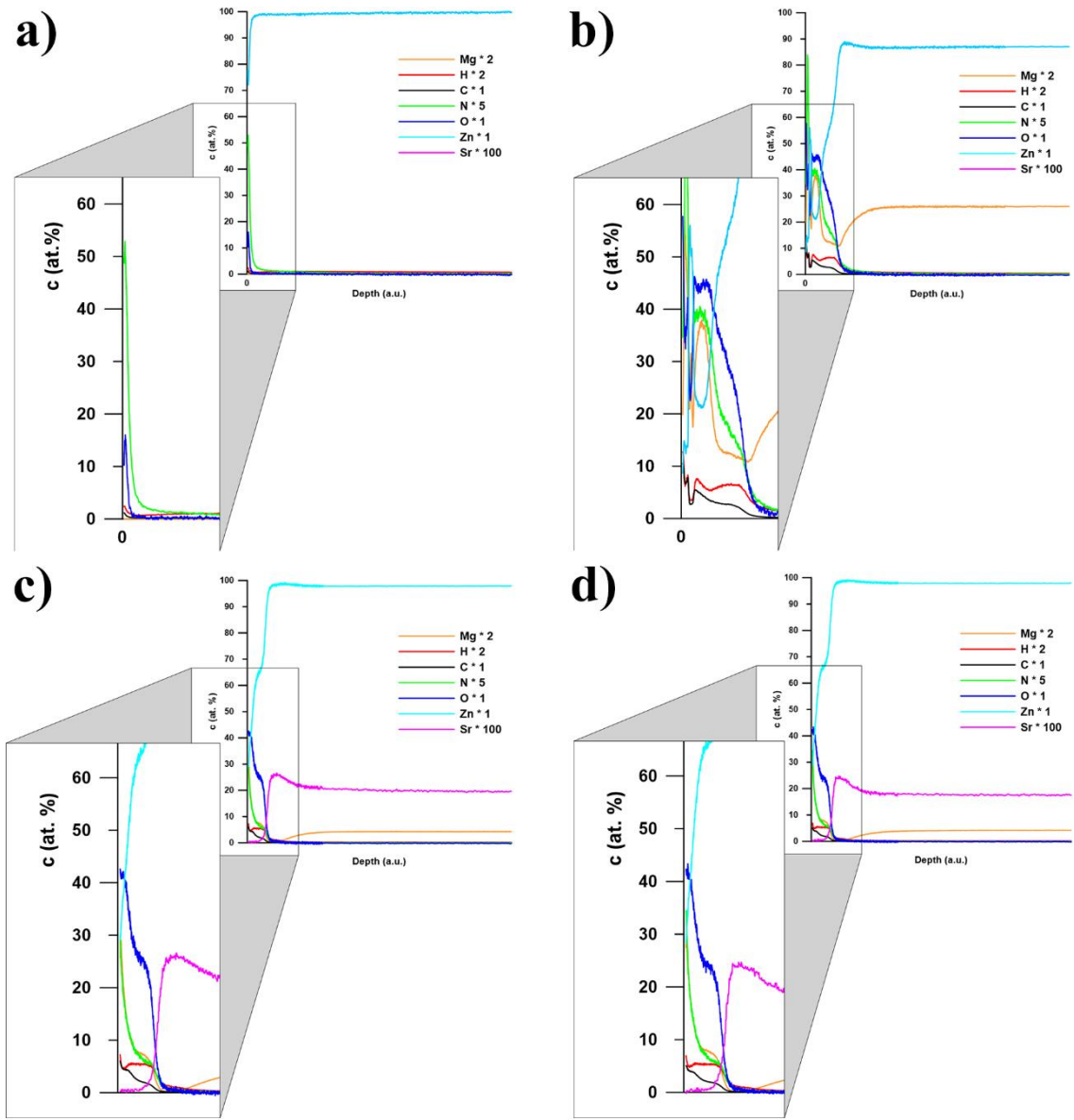

**Fig. 5:** GD-OES depth profiles of a) pure Zn, b) Mg$_2$Zn$_{11}$, c) Zn-0.8Mg-0.2Sr alloy after implantation and d) Zn-0.8Mg-0.2Sr after 3 day of exposure to air

Diffraction patterns acquired in different times are shown in **Fig. 6**. Immediately after the implantation, the minority of the peaks belonging to Mg$_3$N$_2$ phase with a very small intensity were observed (details cut from the diffraction patterns). These peaks were subsequently significantly suppressed in time with no signs of their presence within whole diffraction pattern. Interestingly, other compound containing Mg such as MgO proved by TEM were not found using this method. Besides, disappearance of the Mg$_2$Zn$_{11}$ phase from the affected layer was confirmed after the implantation in all measured times suggesting the phase decomposition during the process of implantation.

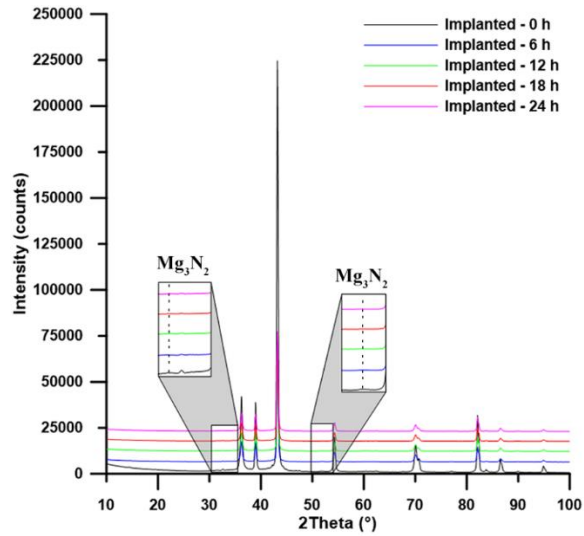

**Fig. 6:** Diffraction patterns obtained during measurements of the phase stability in time

      The topography of the non-implanted and implanted pure Zn, $Mg_2Zn_{11}$ phase, and Zn-0.8Mg-0.2Sr alloy was measured using AFM, and the results are shown in **Fig. 7** along with the average values of the Ra parameter. As seen from the maps acquired on the polished samples, a higher value of the Ra parameter was observed in the case of the implanted $Mg_2Zn_{11}$ phase and Zn-0.8Mg-0.2Sr alloy, where the values reached approximately the same levels. Generally, the roughness of the surface increased after ion implantation within all samples. An interesting observation was made for the implanted $Mg_2Zn_{11}$ phase and Zn-0.8Mg-0.2Sr alloy, where the overall change in height was similar, but the value of the Ra parameter significantly changed. These observations point out a more wrinkled surface (higher density of peaks) in the case of the Zn-0.8Mg-0.2Sr alloy. Additionally, rounded particles were observed on the surface of the implanted alloy (**Fig. 7c**).

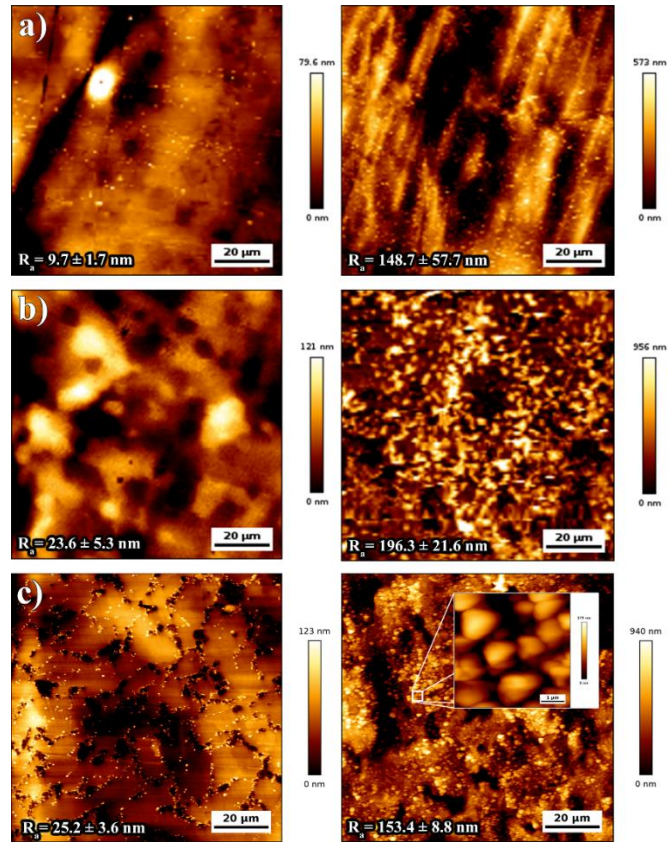

**Fig. 7:** Topography of nonimplanted (left) and implanted (right) a) pure Zn, b) $Mg_2Zn_{11}$ phase, and c) Zn-0.8Mg-0.2Sr alloy with an average values of Ra parameter

## 4. Discussion

It is evident from the results, shown above, that only $Mg_3N_2$ formed during the implantation process. The reason, why $Zn_3N_2$ was not found, could be explained by the thermodynamic stability of Zn and Mg nitrides and the affinity between Zn/Mg and N. The enthalpy of formation is the first crucial indicator of the stability of the compounds. If the formation enthalpy is negative, it means the compound is exothermic and thermodynamically stable with respect to its constituents elements in their standard state. The thermodynamic stability of $Zn_3N_2$ (-22.6 kJ/mol) is quite low due to its negligible formation enthalpy in contrast to $Mg_3N_2$ (-460.7 kJ/mol) [39]. The low formation energy results in the thermodynamic instability of larger zinc nitride crystallites. The possible formation of very small zinc nitride crystals is not excluded, but their size is below the detection limit of diffraction. Therefore, the thermodynamic preference clearly favors magnesium nitride over zinc nitride, as can be seen in **Fig. 8a**. The second crucial factor for formation of nitrides is reaction Gibbs energy. If we consider an implantation temperature of 300 °C, under these conditions, the reaction Gibbs energy for $Zn_3N_2$ is positive (95.668 kJ/mol) while for $Mg_3N_2$ is negative (-345.519 kJ/mol), hence $Zn_3N_2$ does not form spontaneously. The values of Gibbs energy of pure constituents were calculated according to [40]. The third factor affecting the formation of nitrides is phase stability which is presented in phase diagrams. The phase diagram indicates the presence of a magnesium nitride phase, but no stable zinc nitride phase can be found. Although ionic implantation by nitrogen ions may have created some zinc nitride phase, the experimental temperature and the thermodynamic preference for magnesium nitride shift the phase equilibrium clearly towards the magnesium compound. **Fig. 8b** shows the existence of individual phases under the specific conditions for the $Mg-O_2-N_2$ system. Thermodynamic equilibrium enables the coexistence of magnesium oxide and nitride in a low oxygen environment but indicates a shift in composition. This can explain the higher concentration of nitrogen in MgO phase regions compared to zinc regions. This may also elucidate changes in morphology for N-implanted alloys, as the changes are localized in magnesium-rich regions and may be explained by chemical reactions with N ions and subsequent transformation into oxides. The thermodynamic explanation is in good agreement with the obtained results, but signs of active diffusion can also be observed. These indications of diffusion are especially visible in the **Fig. 4**, where a thin layer of Zn is evident under the porous Zn layer or MgO. Clearly, there is a combination of driving forces in the form of diffusion and chemical reaction, making the process even more complex.

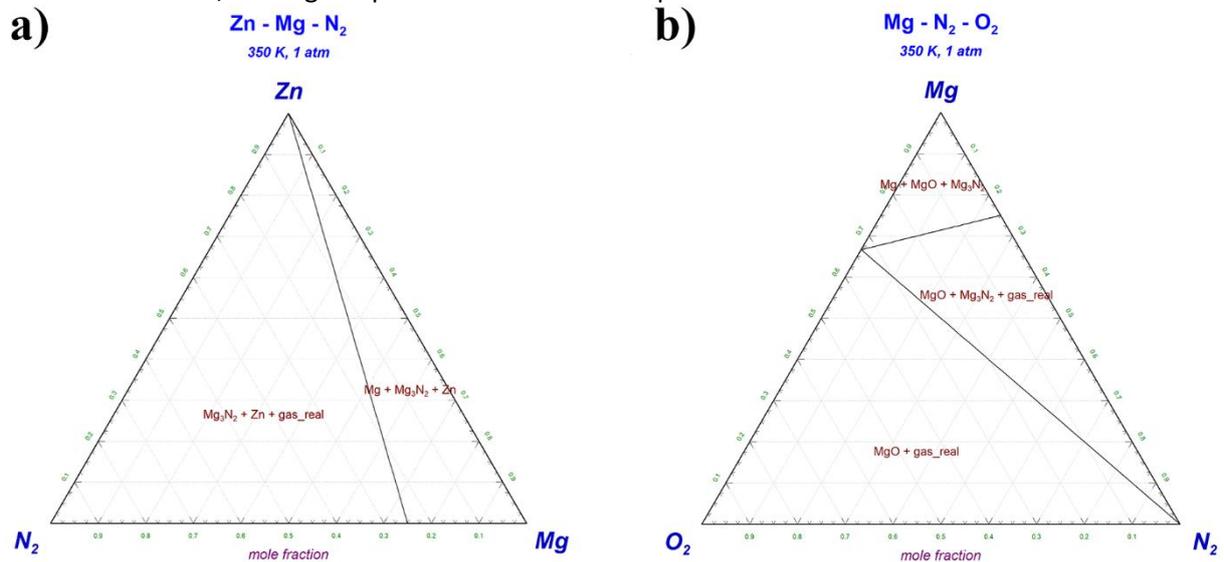

**Fig. 8:** Ternary phase diagrams of a) Zn-Mg-$N_2$ and b) Mg-$N_2$-$O_2$ systems

Determining and quantifying of Mg and Zn nitrides is challenging due to their stability and the limited thickness of the affected layer. These challenges are evident within TEM and XRD analyses. Lamellas were prepared using the FIB technique, creating thin samples with a large area that needed to be transported into the TEM. During transportation, the chemical transformation of $Mg_3N_2$ to MgO can occur, leading to a significant decrease in the nitrides content in the samples. This process can also be active within the bulk sample due to its porous nature and the low thickness of the affected layer. Because of this, understanding the reaction kinetics is crucial and should be determined using various techniques. XRD and GD-OES were employed to confirm the stability of the nitrides. Diffraction patterns were measured over approximately 6 h, suggesting sample exposure to air during measurement, which could introduce potential errors. However, the presence of $Mg_3N_2$ was confirmed using this method only during the first measurement. With prolonged exposure, peaks representing this phase disappeared without being replaced by other Mg-based compounds. This could be explained by the chemical transformation of nitride to oxide. Additionally, the absence of MgO peaks in the diffraction pattern could be due to the overlapping of Zn and MgO peaks and their varying intensities. It was calculated that 63% of the information comes from a depth of 117 nm. Given the overall content and differing intensities of individual peaks, it can be concluded that determining MgO by XRD is not possible in this case (please see the supplementary materials for detailed diffraction patterns). In addition to diffraction patterns, depth profiles were measured immediately after implantation and three days later. These measurements did not confirm any substantial changes in the content of oxygen and nitrogen over time, suggesting the system's stability after three days. Despite obvious contradictions with XRD measurements, it is necessary to mention that the XRD describes only the phase stability while GD-OES only the amount of nitrogen. Given that, the nitrogen should be stored also in the interstitial position in the structure. More precisely, it can be assumed that the content of $Mg_3N_2$ will be minimal in the affected layer after the implantation. With prolonged time (> 6 h), the nitrides transform into oxides. It means that the nitrogen bonded in the nitrides (substitution positions) will not be detected anymore. In addition, other phases containing nitrogen were not detected in the affected layer despite the high content of nitrogen suggesting its presence in the interstitial positions within Zn and MgO. Retention of N ions in these phases could be expected due to enough space to hold nitrogen atoms as interstitials. This is quite common effect resulting from ion implantation [41] and it explains abundance of nitrogen in surface layers.

One of the most important factors is the potential applicability of the modified surfaces within biodegradable implant applications. Despite the absence of further tests toward individual applications in this manuscript, potential directions for future research can be deduced from the obtained results. These directions can be separated into four different groups containing: i) enhancement of cell adhesion and the possibility to perform direct in-vitro testing; ii) creation of nanopore structures allowing the preparation of a base material for a drug delivery system; iii) design of reactive surfaces regulating conditions in the vicinity of the implant; and iv) affecting the degradation rate through different diffusion layers. Prior to further specification of these directions, it is noteworthy that these suggestions are only potential directions and need to be further tested before final approval of their applicability. Despite that, mentioning these possibilities points to the huge potential of ion implantation for biodegradable implant applications, especially for Zn-Mg-based materials.

i) It is well known that problems concerning direct in-vitro testing are common for Zn and its alloys. The reason for this is often connected with the local increase in Zn ion concentration leading to cytotoxic effect. In the case of Zn-Mg alloy without surface modification, the Mg-based phase often disappears in solutions mimicking the body environment due to a more negative potential of the phase compared to pure Zn (galvanic cell), resulting in the dominance of Zn on the surface and cytotoxic behavior. These problems can be effectively solved by the ion implantation process through changes in Zn/Mg ratio and formation of more stable compounds (MgO, $Mg(OH)_2$).

ii) Drug delivery systems are designed to gradually release the drug under specific conditions, such as pH levels, or after a certain amount of time. To achieve these conditions, the materials need to be able to store/attach the drug and release it in the required place. One possible approach is using porous structures as reservoirs for drugs. Porous surfaces of biodegradable Zn alloys created by oversaturation with N could also be used for the preparation of these systems. High fluences of nitrogen ions led to the formation of pore structures on the surface of pure Zn and the $Mg_2Zn_{11}$ phase. However, the dose was insufficient for creating such structures within the alloy, suggesting the need for a further increase in dose to achieve the desired structures.

iii) It was demonstrated in this manuscript that the ion implantation process can lead to the decomposition of existing phases and the creation of a new phase with a different, reactive character. Specifically, the $Mg_2Zn_{11}$ phase decomposes during the process, resulting in the formation of thin layers of MgO and $Mg_3N_2$ compounds. However, these compounds are unstable in a water (body) environment and can lead to the formation of $Mg(OH)_2$ and the release of $NH_3$ through the following reactions:

$$Mg_3N_2 + 3\,H_2O \rightarrow 3\,MgO + 2\,NH_3 \qquad (2)$$

$$MgO + H_2O \rightarrow Mg(OH)_2 \qquad (3)$$

Additionally, these compounds can further dissociate in the body's environment into $Mg^{2+}$ and $OH^-$ ions [42], affecting the pH near the implant. This process could be beneficial in reducing inflammation by increasing the pH around the implanted material. These reactions can be controlled by the composition of the initial material, enhancing the overall performance of the implant shortly after surgery.

iv) As observed in **Fig. 4**, various diffusion layers form on the material surface after the ion implantation process. These layers and their arrangement will play a significant role in the resulting corrosion behavior. By considering the different corrosion rates, the material can be further designed to meet the requirements of the implant based on its function and location.

**Conclusion**

The influence of high fluences ($17\cdot10^{17}$ ions/cm²) of implanted nitrogen ions on the microstructure, topography, elemental, and phase composition of the Zn-0.8Mg-0.2Sr alloy was described, and the most important findings arising from the manuscript can be divided into the following points:

i) High fluences of N ions during the ion implantation process led to the creation of nanopore structures on the surface of pure Zn. This was caused by the oversaturation of the surface with N ions.

ii) In the case of the $Mg_2Zn_{11}$ intermetallic phase, the decomposition of the phase with subsequent diffusion of Mg toward the surface and the creation of a pure Zn layer under the Mg-based layer was observed. By this process, the concentration of Mg in terms of area significantly increased.

iii) Despite the similar behavior of the individual areas in the Zn-0.8Mg-0.2Sr alloy, the oversaturation of these regions was not achieved using the same fluences.

iv) Depth profiles of N concentration point out to the possible accommodation of implanted ions into interstitial positions after 6 h of air exposure.

The aforementioned points support the use of the ion implantation process in the area of hexagonal biodegradable metallic materials by highlighting the beneficial changes that occur during the process. Coupled with the simplicity of the process and the potential affection of wide range of material properties, this technique appears to have significant potential to be used for the surface pretreatment of orthopedic implants.

**Data availability**
The used data are accessible via the Zenodo repository: http://doi.org/10.5281/zenodo.12568954.

**Acknowledgement**
This research was funded by Czech Science Foundation, grant number 23-05592S and Ferroic Multifunctionalities project, supported by the Ministry of Education, Youth, and Sports of the Czech Republic. Project No. CZ.02.01.01/00/22_008/0004591, co-funded by the European Union. CzechNanoLab project LM2023051 funded by MEYS CR is gratefully acknowledged for the financial support of the measurements/sample fabrication at LNSM Research Infrastructure.